\documentclass[aps,superscriptaddress,showpacs,twocolumn]{revtex4-1}
\usepackage[pdftex]{graphicx}
\usepackage{graphicx, epsfig}
\usepackage{amsmath, amssymb}
\usepackage{color}
\usepackage{subfigure}
\usepackage{braket}
\usepackage{fancyhdr}
\usepackage{amsfonts,amsthm,epstopdf}
\usepackage{bbm} 
\theoremstyle{definition}

\theoremstyle{plain}

\newcommand{\Eq}[1]{Eq.(\ref{#1})}
\newcommand{\Sec}[1]{Sec.(\ref{#1})}
\newcommand{\Fig}[1]{Fig.(\ref{#1})}

\newcommand{\be}{\begin{equation}}
\newcommand{\ee}{\end{equation}}
\newcommand{\bea}{\begin{eqnarray}}
\newcommand{\eea}{\end{eqnarray}}
\newcommand{\ba}{\begin{align}}
\newcommand{\ea}{\end{align}}
\newcommand{\SKIP}[1]{}
\def \w{w}
\def \bd {\chi}  

\def \ra{\rightarrow}

\def \eps{\epsilon}

\begin{document}
\title{TensorNetwork on TensorFlow: Entanglement Renormalization for quantum critical lattice models}

\author{Martin Ganahl}
\author{Ashley Milsted}
\affiliation{Perimeter Institute for Theoretical Physics, 31 Caroline Street North, Waterloo, ON N2L 2Y5, Canada}
\author{Stefan Leichenauer}
\author{Jack Hidary}
\affiliation{Alphabet (Google) X, Mountain View, CA 94043, USA}
\author{Guifre Vidal}
\affiliation{Perimeter Institute for Theoretical Physics, 31 Caroline Street North, Waterloo, ON N2L 2Y5, Canada}
\affiliation{Alphabet (Google) X, Mountain View, CA 94043, USA}
\begin{abstract}
  We use TensorNetwork \cite{roberts_tensornetwork:_2019,tensornetworkAPI},
  a recently developed API for performing tensor network contractions using accelerated backends such as TensorFlow
  \cite{tensorflow2015-whitepaper}, to implement
  an optimization algorithm for the Multi-scale Entanglement Renormalization Ansatz (MERA).
  We use the MERA to approximate the ground state wave function of
  the infinite, one-dimensional transverse field Ising model at criticality,
  and extract conformal data from the optimized ansatz. Comparing runtimes of the optimization
  on CPUs vs. GPU, we report a very significant speed-up, up to a factor
  of 200, of the optimization algorithm when run on a GPU.
\end{abstract}

\maketitle

\section{Introduction}
\begin{figure}[!]
  \includegraphics[width=1\columnwidth]{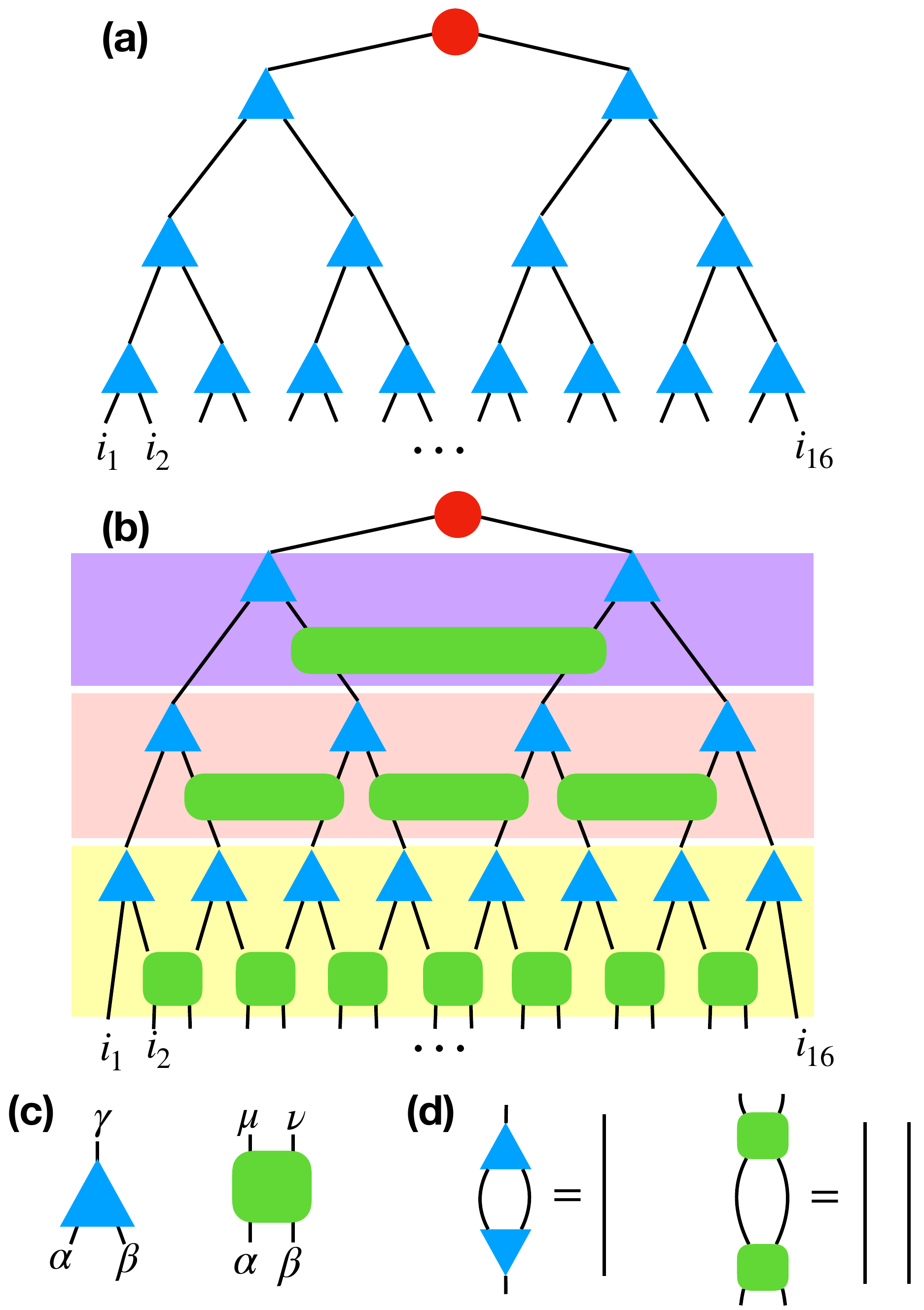}
  \caption{(a) Binary tree tensor network for a wave function $\ket{\Psi}$ of a many-body spin chain
    with 16 sites.  Blue triangles are isometric rank-3 tensors $\w$. The red dot at the top is
    a rank-2 tensor.
    (b) Binary MERA tensor network of a wave function $\ket{\Psi}$ for a spin chain with 16 sites.
    Blue triangles are isometries $\w$, green squares are disentanglers
    $u$. The physical degrees of freedom reside at the bottom of the network and are denoted by
    $i_n$. We have highlighted the different layers of the network in different colors.
    (c) Index ordering of isometries and disentanglers. (d) Isometric and unitary constraints of $\w$ and $u$.
  }\label{fig:MERA}
\end{figure}
Tensor networks are efficient and very powerful parametrizations of
restricted classes of vectors in high-dimensional
vector spaces. They were originally developed to
simulate many-body quantum systems in condensed matter physics.
Tensor networks like the Matrix Product State (MPS) \cite{fannes_finitely_1992,white_density_1992,bridgeman_hand-waving_2017},
the Multi-Scale Entanglement Renormalization Ansatz (MERA) \cite{vidal_entanglement_2007,vidal_class_2006} 
or Projected Entangled Pair States (PEPS) \cite{verstraete_criticality_2006,verstraete_renormalization_2004}
are nowadays routinely used to approximate ground states of quantum many-body Hamiltonians in one (1d)
and two (2d) spatial dimensions or to simulate real-time dynamics of many-body quantum states.
Over the past decade it has become evident that tensor networks are also useful well outside their original
scope, and they have for example found applications in quantum-chemistry \cite{white_ab_1999,chan_introduction_2007,szalay_tensor_2015,
  krumnow_fermionic_2016, white_multi-sliced_2019}, real material calculations
\cite{bauernfeind_fork_2017, ganahl_efficient_2015, ganahl_chebyshev_2014},
quantum field theory \cite{verstraete_continuous_2010, haegeman_entanglement_2013,ganahl_continuous_2017},
machine learning \cite{stoudenmire_supervised_2016,stoudenmire_learning_2018,evenbly_number-state_2019} and
even quantum gravity \cite{swingle_entanglement_2012,beny_causal_2013,czech_tensor_2016,
  bao_sitter_2017,milsted_geometric_2018}. 

TensorFlow \cite{tensorflow2015-whitepaper} is a free, open source software library
for data flow and differentiable programming, developed
by the Google Brain team, that can be used for a range
of tasks including machine learning applications such as
neural networks. Recently, the open source library TensorNetwork
\cite{roberts_tensornetwork:_2019,tensornetworkAPI} has been released to facilitate the development and implementation
of tensor network algorithms in TensorFlow.

In previous work  we have provided benchmark
results for the use of TensorNetwork for tensor network computations in condensed matter
physics \cite{milsted_tensornetwork_2019} and machine learning \cite{efthymiou_tensornetwork_2019}.
In \cite{milsted_tensornetwork_2019} we have used TensorNetwork to implement and benchmark an optimization algorithm
for so-called Tree Tensor Networks (TTNs) to approximate
the ground state of the Ising model on a torus. Using TensorNetwork's TensorFlow backend to run the optimization
on accelerated hardware, we obtained speed-ups of a factor of 100 over optimization on CPUs.
Similar speed-ups are reported in \cite{efthymiou_tensornetwork_2019} for applications
of TensorNetwork to classification tasks in the area of machine learning.
\begin{figure}[!]
  \includegraphics[width=1\columnwidth]{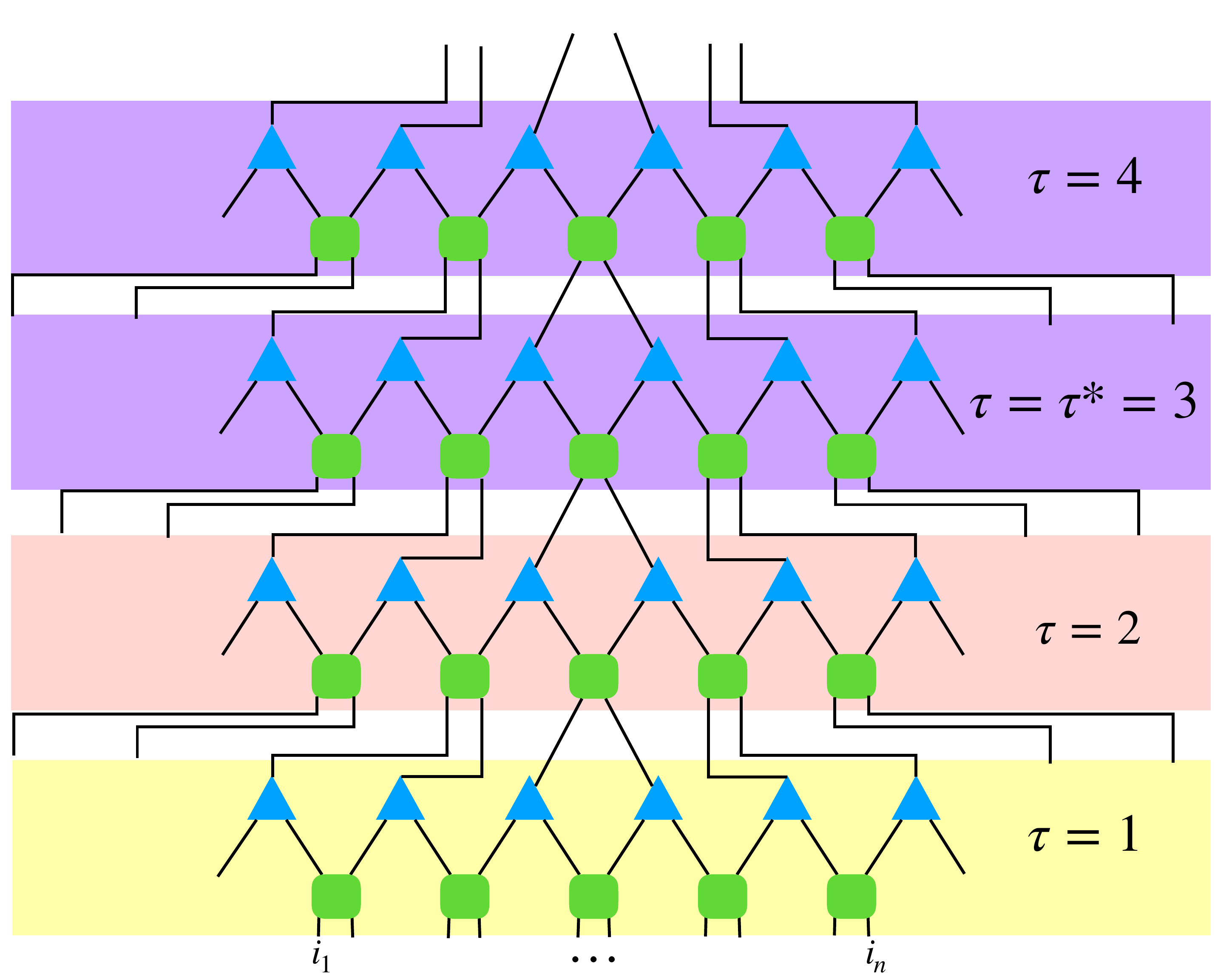}
  \caption{First four layers of a scale invariant binary MERA. The transitional layers $\tau=1$ and $\tau=2$
    are highlighted in yellow and red, scale invariant layers $\tau\geq 3$ are highlighted in purple.
    Tensors in the scale invariant layers $\tau\geq 3$ are all identical, whereas tensors in the transitional
    layers differ from layer to layer.}\label{fig:SI_MERA}
\end{figure}

In this manuscript we use TensorNetwork to implement an optimization algorithm for MERA.
The purpose of this paper is to (1) provide sample code which shows
how to use TensorNetwork to implement a MERA optimization. This is of interest to tensor network
practitioners who want to get started quickly with TensorNetwork, as well as for newcomers who want
to understand basic concepts of MERA optimizations; (2) to benchmark runtimes for MERA on CPU and GPU,
and analyze in detail how the computational cost is distributed among different computational tasks.

To motivate the MERA, let us briefly review the basics of binary TTNs, shown in \Fig{fig:MERA} (a).
In a TTN all tensors (shown as blue triangles) are structurally identical (apart from the top-most tensor, shown in red).
The TTN shown in \Fig{fig:MERA} (a) can be contracted and optimized efficiently.
Similar to a TTN, the MERA (shown in \Fig{fig:MERA} (b)) is a tensor network designed to approximate ground states of
many-body Hamiltonians. 
The MERA is a powerful generalization of a TTN which introduces a 
second set of structurally different tensors, called {\it disentanglers} (shown as green squares in \Fig{fig:MERA} (b))
into the network. Abundant numerical evidence has long established that the MERA
is well suited to approximating ground states of critical quantum systems \cite{evenbly_algorithms_2009},
more so than for example the simple TTN in \Fig{fig:MERA} (a).
In this paper we review a standard MERA energy minimization algorithm from Refs. \cite{evenbly_algorithms_2009,pfeifer_entanglement_2009,evenbly_quantum_2011},
which results in a (scale invariant) MERA representation of an infinite, critical quantum spin chain.
While MERA is a more powerful tensor network than a TTN, the addition of disentanglers
significantly increases the computational complexity for fixed bond dimension $\chi$
(the leading cost changes from $\chi^4$ to $\chi^9$).

It is expected that, similar to TTNs \cite{milsted_tensornetwork_2019},
MERA algorithms can be considerably sped up by running them on accelerated hardware
like GPUs or TPUs. Using TensorNetwork with TensorFlow as backend, we confirm this expectation and show that running
a MERA optimization algorithm on state-of-the-art accelerated hardware can yield speed-ups of up to
a factor of 200 as compared to running it on a CPU.
The code for all calculations can be downloaded from \cite{tensornetworkAPI}.

\section{Review of the Multi-scale Entanglement Renormalization Ansatz}
The purpose of this section is to introduce the reader to notation and give a
high-level introduction to MERA. For a more comprehensive review, we refer the reader to
\cite{evenbly_algorithms_2009}.
The MERA is a class of variational tensor network wave functions which can be used to efficiently approximate
ground states of critical quantum many-body Hamiltonians.
In the following we will use the {\it scale invariant} MERA \cite{pfeifer_entanglement_2009}
(see \Sec{sec:si_mera} and \Fig{fig:SI_MERA}), to approximate
the ground state of critical quantum lattice models in the thermodynamic limit on a
one dimensional (1d) lattice $\mathcal{L}$.
As a benchmark example, we will consider the quantum Ising model with Hamiltonian
\begin{align}
  H = \sum_{n\in \mathbbm{Z}} -X_n X_{n+1} - h Z_n \equiv \sum_{n\in\mathbbm{Z}} h_{n,n+1}\label{eq:ham}
\end{align}
with $X = \left(
\begin{array}{cc}
  0 & 1\\
  1& 0
\end{array} \right)
$
and
$Z = 
\left(
\begin{array}{cc}
  1 & 0\\
  0& -1
\end{array} \right)
$
Pauli spin operators, i.e. each site on the lattice $\mathcal{L}$ hosts a spin $1/2$ degree of freedom.
The Hamiltonian has a phase transition at a magnetic field $h=1$ from a paramagnetic ($h>1$) to a ferromagnetic
($h<1$) phase. At the critical point $h=1$ the spectrum of $H$ becomes gapless and ground state correlations decay
algebraically (that is, as a power law $x^{-2\Delta}$ with distance $x$).
One of the key advantages of MERA over other
variational classes of wave functions is its ability to {\it exactly} capture algebraic correlations at
arbitrary distances. This makes it an ideal ansatz for ground states of critical quantum systems.

\begin{figure}[!]
  \includegraphics[width=1.0\columnwidth]{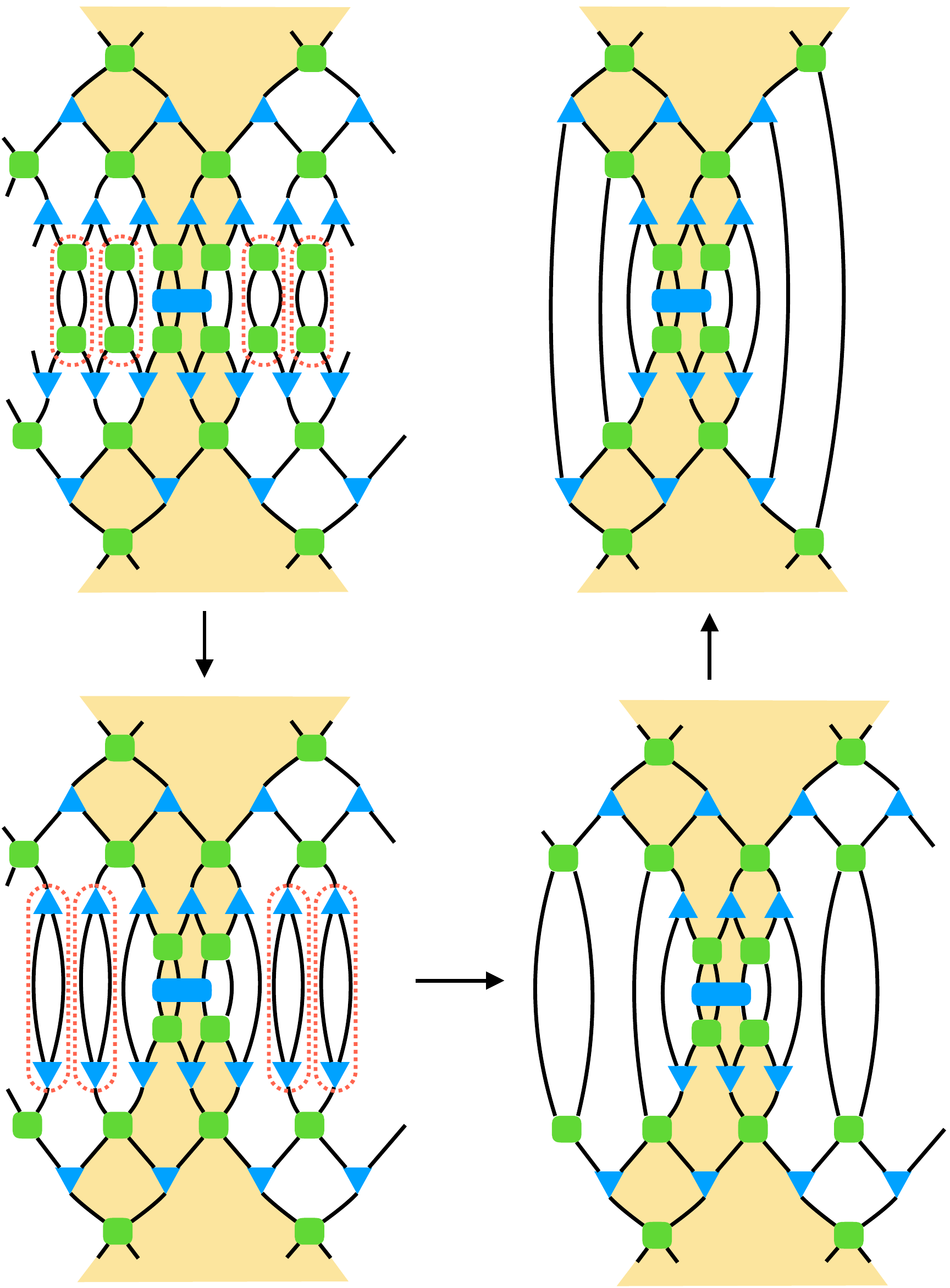}
  \caption{Computation of the expectation value $\braket{\Psi|h^{\tau=1}_{n, n+1, n+2}|\Psi}$ of a 
    a local observable $h^{\tau=1}_{n,n+1, n+2}$ for a scale invariant binary MERA. In order to calculate
    $\braket{\Psi|h^{\tau=1}_{n, n+1, n+2}|\Psi}$ we need to contract the infinite tensor network shown in the
    top left corner. Highlighted in orange is the {\it past causal cone} of the sites $n, n+1, n+2$.
    The past causal cone contains all
    tensors which can affect the sites $n, n+1, n+2$. Due to the isometric and unitary constraints
    \Eq{eq:iso} and \Eq{eq:unitarity}, many of the tensor contractions in the network become
    trivial. Starting from the top left, tensors inside the red dashed ellipses are sequentially
    replaced with trivial contractions. The final network contains only tensors inside the past causal
    cone of sites $n, n+1$ and $n+2$.}
  \label{fig:MERA_causal}
\end{figure}

\subsection{Scale invariant MERA}\label{sec:si_mera}
The scale invariant MERA is a tensor network representation of a many-body wave function $\ket{\Psi}$
on an {\it infinite} lattice $\mathcal{L}$. \Fig{fig:SI_MERA} shows an example of a 
scale invariant MERA. The physical degrees of freedom
of this lattice are denoted by $i_n$, where $n \in \mathbbm{Z}$
denotes the lattice site, and $i_n \in \{1,2\}$
for a two level quantum spin degree of freedom, such as in the transverse field Ising model.
In this paper we will focus on the so-called {\it binary} MERA. It
consists of two types of tensors (see \Fig{fig:MERA} (b) and (c)) called 
isometries $\w$ and disentanglers $u$. Isometries $[\w]_{\alpha\beta}^{\gamma}$ are rank-3 tensors
of dimension $\bd\times \bd\times \bd$ which obey the isometric constraint
\begin{align}
  \sum_{\alpha\beta}[\w]_{\alpha\beta}^{\gamma}[\overline{\w}]_{\alpha\beta}^{\gamma'} = \delta_{\gamma\gamma'}\label{eq:iso}
\end{align}
while disentanglers $[u]_{\alpha\beta}^{\mu\nu}$ are rank-4 tensors of dimension $\bd \times \bd\times \bd\times \bd$ obeying
\begin{align}\label{eq:unitarity}
  \sum_{\alpha\beta}[u]_{\alpha\beta}^{\mu\nu}[\overline{u}]_{\alpha\beta}^{\mu'\nu'} = \delta_{\mu\mu'}\delta_{\nu\nu'}
\end{align}
see \Fig{fig:MERA} (d). Square brackets are used to denote tensor elements, overlines denote complex conjugation,
and $\delta_{\mu\nu}$ is the Kronecker delta.
To simplify explanations, we assume that all isometries $\w$ and disentanglers $u$ have identical dimensions
$\bd$.
A MERA is organized into {\it layers $\tau = 1, 2, 3, \dots\;$}. 
In \Fig{fig:SI_MERA} we show the first four layers of a scale invariant MERA.
Each layer corresponds to a different {\it scale} \cite{evenbly_algorithms_2009} of the network, 
and contains a row of isometries $\w_{\tau n}$ and a row of disentanglers $u_{\tau n}$,
where $\tau$ and $n$ label layer (or scale) and position, respectively.
In the following we impose translational invariance by choosing $\w_{\tau n}$ and $u_{\tau n}$ to
be independent of $n$ within each layer $\tau$, and will henceforth omit the subscript $n$.
The physical degrees of freedom $i_n$ reside at the bottom of the network.
The scale invariant MERA consists of an infinite number of layers $\tau$. The layers are divided into
{\it transitional layers} with $\tau < \tau^*$ (where $\tau^*$ is a hyper parameter of the network)
and {\it scale invariant layers} $\tau \geq \tau^*$.
For instance, in \Fig{fig:SI_MERA} we have $\tau^*=3$, indicating that there are two transitional
layers of tensors for $\tau=1,2$ before the scale invariant layers for $\tau\geq 3$.
For the scale invariant layers, all tensors $u_{\tau}, \w_{\tau}$
are identical and we will occasionally denote them by $\w$ and $u$ without subscript.
Tensors in the transitional layers are different for each layer.
That is, if $\tau^*=3$ then the entire MERA is specified by the six tensors
$u_1, \w_1, u_2, \w_2, u_3, \w_3$.

\subsection{Causal cone, local observables, scaling operators}
\begin{figure}
  \includegraphics[width=1.0\columnwidth]{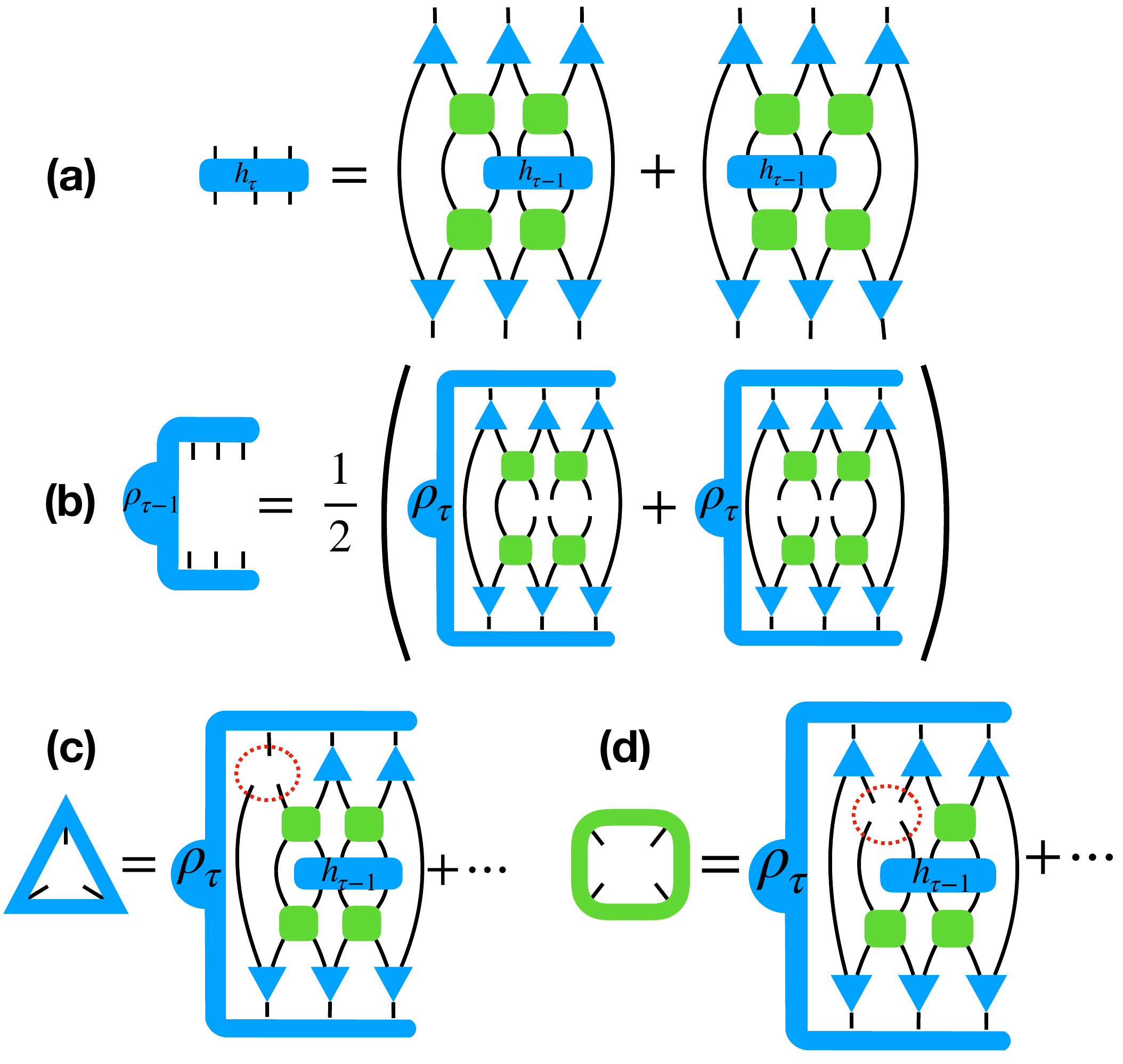}
  \caption{(a) Definition of the ascending super-operator $\mathcal{A}$. The ascending super-operator $\mathcal{A}$
    ascends a local operator, represented as a blue rectangle with six legs,
    up by one layer of the MERA. As an example, we show the ascending operation of the Hamiltonian density.
    Note that operators with support on three sites
    are mapped into operators with support on three sites on a coarser lattice.
    (b) Definition of the descending super-operator $\mathcal{D}$. The descending super-operator
    $\mathcal{D}$ descends a reduced density matrix $\rho_{\tau}$, represented by a blue object with six legs, down
    by one layer of the MERA. Reduced density matrices with support on three sites are mapped
    into reduced density matrices with support on three sites on a finer lattice.
    (c) Environment of the isometry $\w_{\tau}$, showing one of six contributions.
    (d) Environment of the disentangler $u_{\tau}$, showing one of four contributions.
    Environment tensors are drawn as a triangle or a square with the tensor legs pointing inward.}\label{fig:MERA_diagrams}  
\end{figure}
In the following we explain how to efficiently compute observables for a given MERA. As
a concrete example, we will focus on the expectation value of the Hamiltonian \Eq{eq:ham}.
In a binary MERA, Hamiltonians appear naturally as sums of three body terms.
It is therefore convenient to regroup the Hamiltonian \Eq{eq:ham} into a sum
of three body terms:
\begin{align}
  h_{n, n+1, n+2}^{\tau=1}&\equiv \frac{1}{2}(h_{n,n+1}\otimes\mathbbm{1}_{n+2} + \mathbbm{1}_{n}\otimes h_{n+1,n+2}).\nonumber\\
  H &= \sum_nh_{n, n+1, n+2}^{\tau=1}\nonumber.
\end{align}
We refer to $h_{n,n+1,n+2}^{\tau=1}$ as the Hamiltonian density in the first layer.
To calculate the expectation value $\bra{\Psi}h_{n, n+1, n+2}^{\tau = 1}\ket{\Psi}$ of a local Hamiltonian term, we need to
contract an infinite network of isometries and disentanglers. \Fig{fig:MERA_causal} shows a diagrammatic
representation of $\bra{\Psi}h_{n, n+1, n+2}^{\tau = 1}\ket{\Psi}$. One of the key features
of the MERA is the presence of a {\it causal cone} \cite{evenbly_algorithms_2009,evenbly_quantum_2011,tensors.net}
which allows for an efficient contraction
of the network. The causal cone for a binary MERA is shown in \Fig{fig:MERA_causal} as highlighted region.
Using the isometric and unitary conditions \Eq{eq:iso} and \Eq{eq:unitarity} (see also \Fig{fig:MERA} (c) and (d)), 
all contractions outside the causal cone can be carried out trivially, 
and the only contractions that need to be performed explicitly are over tensors inside the causal cone.
This is illustrated in \Fig{fig:MERA_causal}.
The contraction of the network shown in the top-right corner of
\Fig{fig:MERA_causal} is done in several steps:
the first step is the contraction of the infinite number of tensors inside the
scale invariant layers of the causal cone. This is achieved by calculating the dominant eigenvector $\rho_{\tau^*}$ of the
{\it descending super operator} $\mathcal{D}$ (defined in
\Fig{fig:MERA_diagrams} (b)) by iterative methods \cite{evenbly_algorithms_2009}.
In the second step, $\rho_{\tau^*}$ is then descended through the transitional layers of the MERA
down to the bottom of the network using the descending super-operator of the transitional layers.
Finally, at the bottom of the network, $\rho_{\tau=1}$ is contracted with the operator $h_{n, n+1, n+2}^{\tau=1}$
to give its expected value.
For the scale invariant binary MERA, all contractions can be carried out at a leading
computational cost of $\bd^9$.

Two hallmarks of quantum criticality are the phenomena
of {\it scale invariance} and {\it universality}.
Scale invariance refers to the fact that at a critical point,
a physical system looks similar when observed at different length scales.
Universality refers to the fact that microscopically different systems
may have identical properties at their respective critical points.
More generally, critical systems can be divided into universality classes.
The universal physical properties of systems within one such class
can be described by so-called {\it scaling operators}. Scaling operators are operators
which transform covariantly under changes of scale. For example, under a
change of scale by a factor of two, a scaling operator $\mathcal{O}$
transforms as \cite{pfeifer_entanglement_2009}
\begin{align}
  \mathcal{O} \ra 2^{-\Delta_{\mathcal{O}}}\mathcal{O}
\end{align}
where $\Delta_{\mathcal{O}}$ is called the scaling dimension of operator
$\mathcal{O}$.
In a so-called {\it conformal field theories}
\cite{di_francesco_conformal_1997}, scaling operators are organized
into distinct {\it conformal towers}. Operators at the bottom of a tower
are called {\it primaries}, and those higher up are called {\it descendants}.
There is one tower for each primary operator. The number of primary operators
is a property of the physical system.
For the Ising model at criticality, the low energy spectrum of \Eq{eq:ham}
is described by the so-called {\it Ising conformal field theory} (Ising CFT) \cite{di_francesco_conformal_1997}
with {\it central charge}
$c=\frac{1}{2}$ and three primary operators $\mathbbm{1}, \sigma$ and $\eps$, with scaling dimensions
$\Delta_{\mathbbm{1}}=0,\Delta_{\sigma}=\frac{1}{8}$ and $\Delta_{\eps}=1$, respectively.

One of the highlights of MERA is its realization of a discrete scale transformation by a factor of two for the binary MERA
(we refer the reader to \cite{evenbly_algorithms_2009,pfeifer_entanglement_2009,evenbly_quantum_2011,tensors.net} for details).
From an optimized MERA wave function, lattice versions of
scaling operators and their corresponding scaling dimensions can be
approximately obtained from diagonalizing the scale invariant
ascending super-operator $\mathcal{A}$ of \Fig{fig:MERA_diagrams} (a).
\begin{figure}
  \includegraphics[width=1.0\columnwidth]{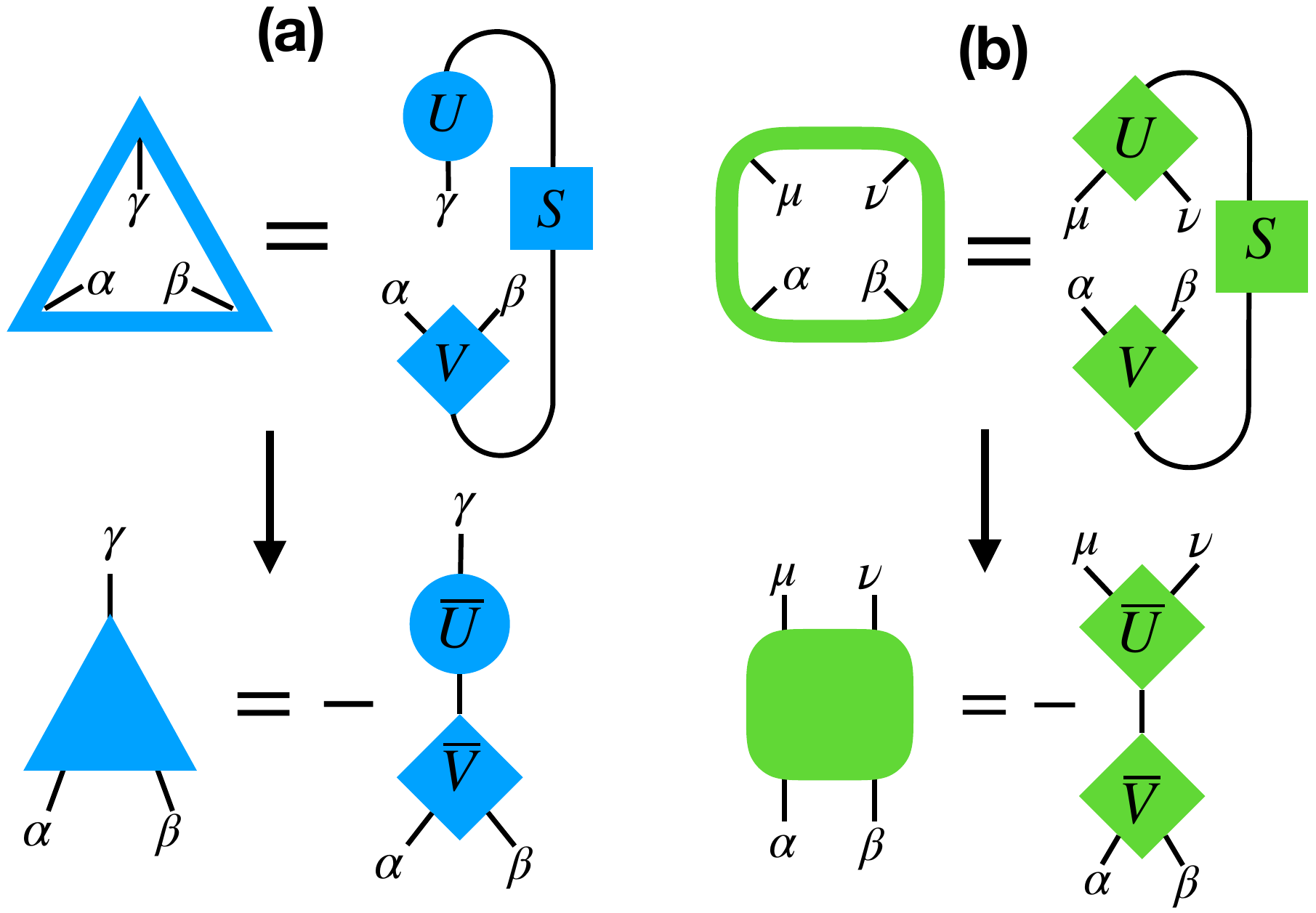}
  \caption{Update to the isometries $\w$ (a) and disentanglers $u$ (b). First we SVD the environments
    to obtain the tensors $U,V$ by appropriate blocking of indices.
    The new isometry or disentangler are given by contracting
    $\overline U$ and $\overline V$ as shown in the bottom parts of the figure.}\label{fig:update}  
\end{figure}

\subsection{Optimization}
We use a standard variational optimization algorithm to optimize the scale invariant MERA
to approximate the ground state of Hamiltonian $H$ in \Eq{eq:ham}.
The algorithm iteratively updates isometries and disentanglers to lower the expectation value of the energy,
which is calculated as outlined in the previous section.
We call one round of updates a {\it sweep}.
One sweep consists of several steps, which we will describe in the following.
In the first step, the isometries and disentanglers in
the scale invariant layers $\tau\geq \tau^*$
are updated. In the second step, isometries and disentanglers
in the transitional layers $\tau<\tau^*$ are updated
layer by layer (we refer the reader to \cite{evenbly_algorithms_2009,evenbly_quantum_2011,tensors.net} for
a detailed discussion of optimization algorithms).
To update isometries $\w_{\tau}$ and disentanglers $u_{\tau}$ in a layer $\tau$, we need to calculate the {\it environments}
of the corresponding tensors. The environments of $\w_{\tau}$ and $u_{\tau}$ are rank-3 and rank-4 tensors, respectively.
Exemplary contributions to the environments of the isometry $\w_{\tau}$ and the disentangler $u_{\tau}$ are shown in
\Fig{fig:MERA_diagrams} (c) and (d).
To calculate the environments in layer $\tau$, we use the descending super-operator $\mathcal{D}$ to descend $\rho_{\tau^*}$ to
layer $\tau + 1$, and the ascending super-operator $\mathcal{A}$ to ascend the local Hamiltonian density
$h_{n, n+1, n+2}^{\tau=1}$ up to layer $\tau$ .
The full environments consist of sums of several contributions of similar form (see \Fig{fig:all_envs}).
Finally, the updates to $u_{\tau}$ and $\w_{\tau}$ are obtained from an SVD of the environments.
This is shown graphically in \Fig{fig:update}. The sweeps are repeated until sufficient convergence is reached.

\section{Benchmark results}
In the following we present benchmark results for the MERA approximation of the ground state
of the critical transverse field Ising model at magnetic field $h=1$, \Eq{eq:ham}.
We have performed calculations for bond dimensions $\bd=4,6,8,10,12, 14$ and $16$.
\subsection{Ground state energy, scaling dimensions}
\begin{figure}
  \includegraphics[width=1\columnwidth]{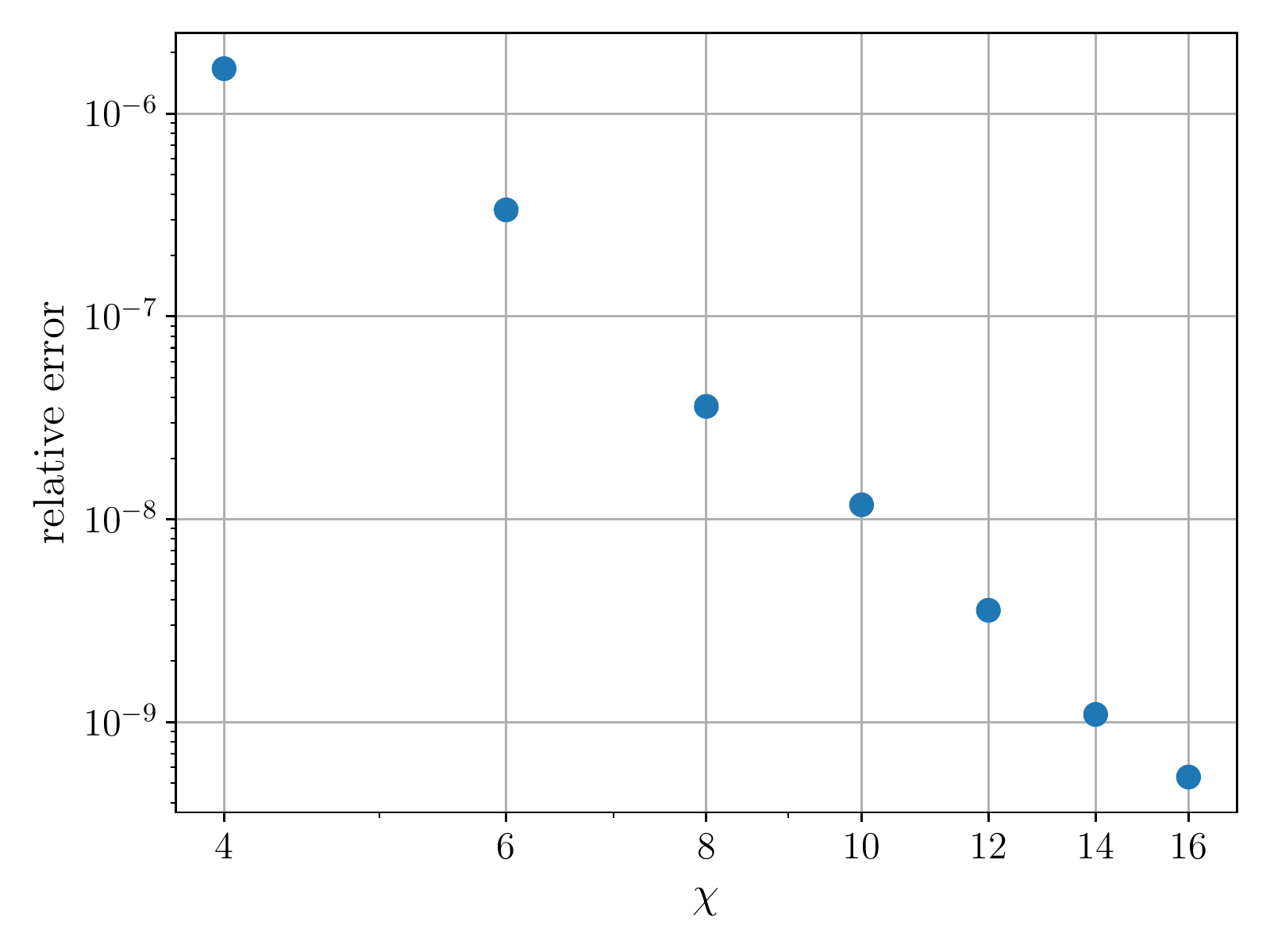}
  \caption{Relative error of the energy of an optimized, scale invariant binary MERA as a function
    of bond dimension $\chi$.}\label{fig:energy}
\end{figure}

In \Fig{fig:energy} we present our benchmark results for the ground state energy
for different bond dimensions $\bd$ of the MERA. The exact ground state energy density is given
by $e_{exact} = -\frac{4}{\pi}$. The relative error in the ground state energy decays
quickly with increasing bond dimension. At our largest bond dimension $\bd=16$,
the error is $\approx 5\times 10^{-10}$.

\begin{figure}
  \includegraphics[width=1\columnwidth]{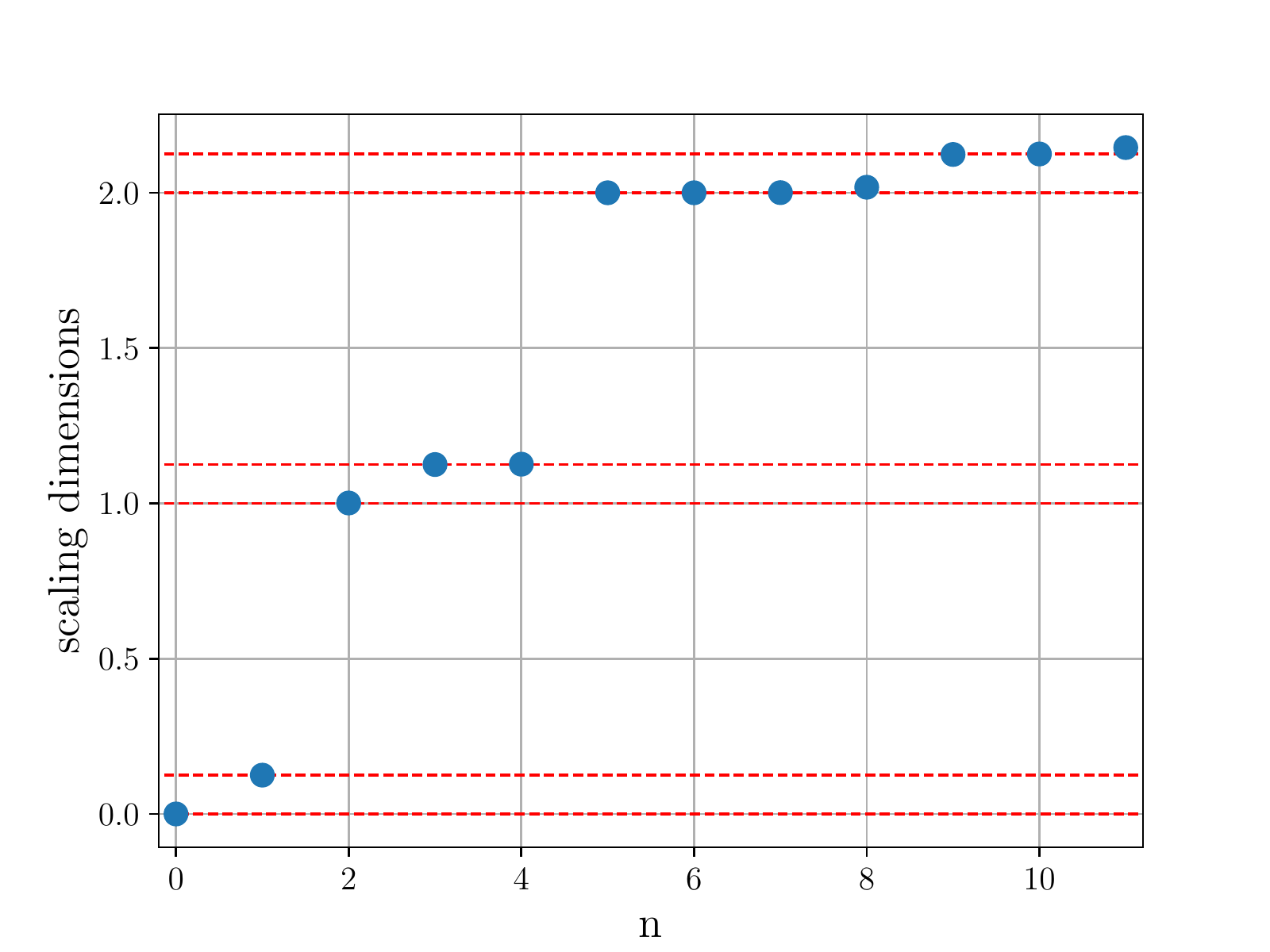}
  \caption{Lowest twelve scaling dimensions (blue dots) obtained from
    the lowest twelve eigenvalues $\lambda_n = 2^{-\Delta_n}$ of the scale invariant ascending super-operator $\mathcal{A}$ ($\chi=16$).
    We show the exact CFT results for the scaling dimensions of the Ising CFT
    as red lines \cite{di_francesco_conformal_1997}.}\label{fig:scaling_dims}  
\end{figure}

In \Fig{fig:scaling_dims} we show the lowest twelve scaling dimensions $\Delta$ obtained
from the eigenvalues $\lambda = 2^{-\Delta}$ of the ascending super-operator $\mathcal{A}$ (blue dots),
together with the exact scaling dimensions of the Ising CFT (red lines), for a bond dimension of
$\bd=16$. We observe a very good agreement with the theoretical results. Larger scaling dimensions
are significantly less accurate.

\subsection{Runtimes}
\begin{figure}
  \includegraphics[width=1\columnwidth]{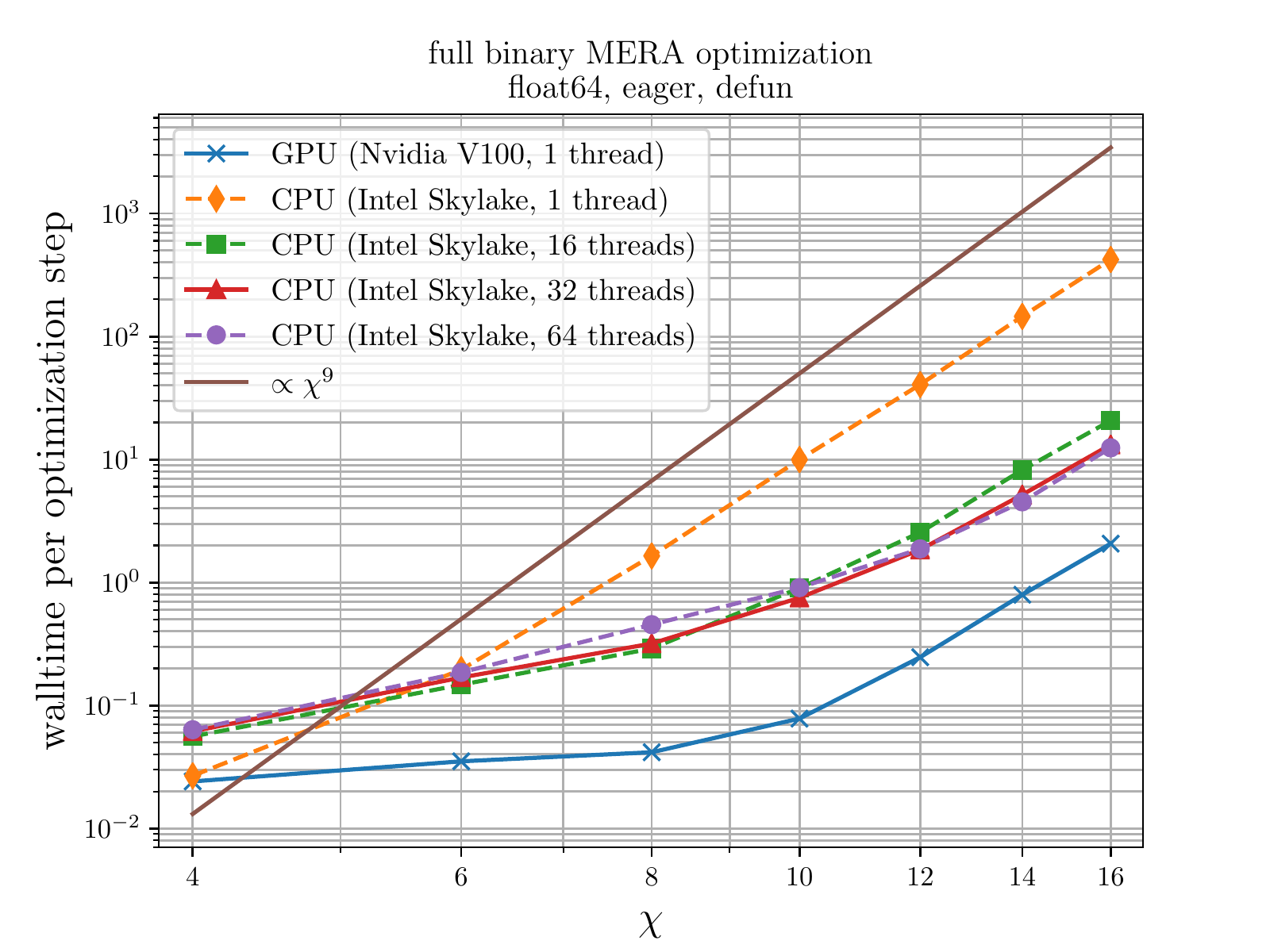}
  \caption{Comparison of average runtimes per optimization step of a scale invariant binary MERA
    for different bond dimensions $\bd$. The asymptotic scaling of the optimization algorithm is $\bd^9$
    (solid brown line). For single-threaded optimization on 1 CPU, we observe a significant speed-up of $200$
    when running the optimization on GPU. Running the optimization on 16 and 32 CPUs with shared memory
    reduces the gap to $6$. Going beyond 32 CPU does not give any additional speed-up.
  }\label{fig:walltimes}    
\end{figure}
\begin{figure*}
  \includegraphics[width=2\columnwidth]{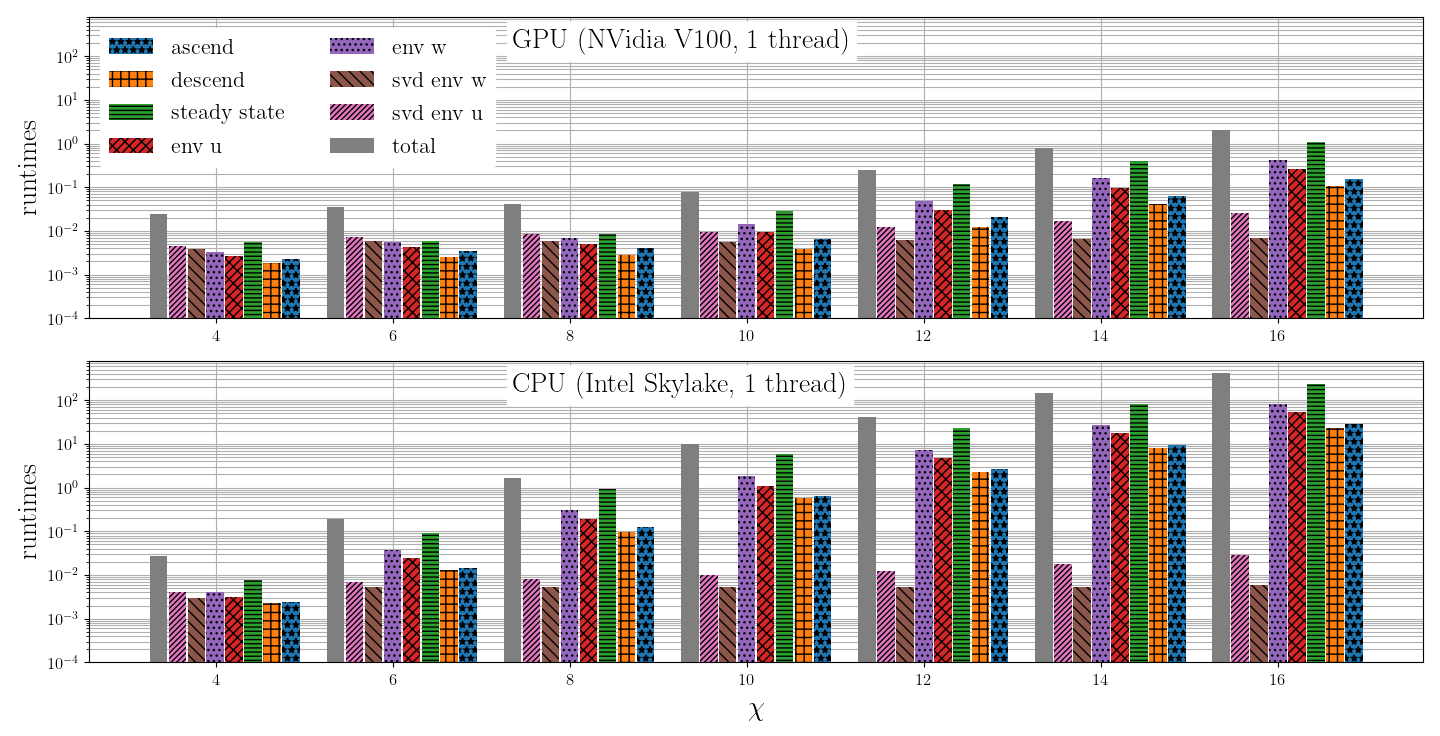}  
  \caption{Runtimes of individual steps of the MERA optimization using single threaded operation.
    The top panel shows results for optimization on GPU, the bottom panel for optimization on CPU.
  }\label{fig:MERAruntimes}
\end{figure*}

One of the key advantages of the TensorNetwork API is the use of TensorFlow as a backend
to perform tensor contractions. Thus, it can 
be easily run on accelerated hardware like GPUs or TPUs, with the benefit
of considerable reduction of runtimes. In the following we present benchmark results
comparing optimization runtimes on CPU and GPU.
We ran our simulations on TensorFlow v1.13.1 built with the Intel math kernel
library (MKL) on Google's cloud compute engine. CPU benchmarks were run on an Intel Skylake
architecture with 1, 16, 32, and 64 cores. The GPU benchmarks were run on an NVIDIA Tesla V100
graphics processor \footnote{SVDs were computed on the host CPU, which was seen to be faster
  than performing the SVD on GPU. Since the SVDs contribute only a small fraction to the total runtime,
  the overall runtimes were essentially unaffected by this choice.}.

The most expensive operations of the optimization algorithm described above scale
asymptotically as $O(\chi^9)$.
In \Fig{fig:walltimes} we compare average runtimes
of one step of the optimization (averaged over twentyx iteration steps) run on CPUs and GPU.
When run on one CPU (single thread), the asymptotic scaling is seen to be $\bd^9$, consistent
with the theoretical expectation.
For single-threaded operation and large bond dimensions, we observe a $200$
fold speed-up of the optimization when run on a GPU,
as compared to CPU. For multi-threaded optimization on multiple CPUs,
we observe a convergence of the speed-up with the number of
CPUs at 32 threads. Compared to 32 CPUs, the GPU is still faster by a factor of $6$.

In \Fig{fig:MERAruntimes} we show the runtimes of the individual steps of a MERA
optimization. The upper panel shows runtimes on a GPU running with a single thread, the
lower panel results for CPU running with a single thread. As expected, in either case
the dominant cost is given by the computation of the steady-state reduced density matrix $\rho^*$ and
the computation of the environments. Running on GPU gives speed-ups of these operations by a factor of ~200.
Note that the SVDs in both cases were performed on CPU. At the largest bond dimension $\chi=16$
the cost for calculating $\rho^*$ is still about 40 times larger than the SVD of the environment of $u$.
This is in contrast to the case of the TTN \cite{milsted_tensornetwork_2019},
where the factor between the most expensive tensor contractions
and the SVD is on the order of 3. For this reason, the MERA optimization algorithm exhibits a greater speed-up
as compared to a TTN optimization when run on a single thread.

\section{Conclusion}
We have used TensorNetwork, an API for tensor contractions with a TensorFlow backend,
to implement the scale invariant MERA optimization algorithm
of Ref. \cite{pfeifer_entanglement_2009, evenbly_quantum_2011}.
As a benchmark, we have used the algorithm to approximate
the ground state of the critical transverse field Ising model in the thermodynamic limit, using
MERA with bond dimensions of up to $\chi = 16$. From the optimized MERA, we calculated
the lowest twelve scaling dimensions of the Ising model, which are
found to be in excellent agreement with their theoretically predicted values.
When run on a GPU, we observe a $200$ fold speed-up as compared to 1 CPU core, and a
speed-up of $6$ when compared to 32 CPUs.

{\bf Acknowledgements} M. Ganahl, A. Milsted and
G. Vidal thank X for their hospitality. X is formerly
known as Google[x] and is part of the Alphabet family of
companies, which includes Google, Verily, Waymo, and
others (www.x.company).
G. Vidal is a CIFAR fellow in the Quantum Information Science Program.
Research at Perimeter Institute is supported by the Government of Canada through
the Department of Innovation, Science and Economic
Development Canada and by the Province of Ontario
through the Ministry of Research, Innovation and Science.

\setcounter{equation}{0}
\setcounter{figure}{0}
\setcounter{table}{0}
\renewcommand{\theequation}{A\arabic{equation}}
\renewcommand{\thefigure}{A\arabic{figure}}
\begin{figure*}[!]
  \includegraphics[width=1.5\columnwidth]{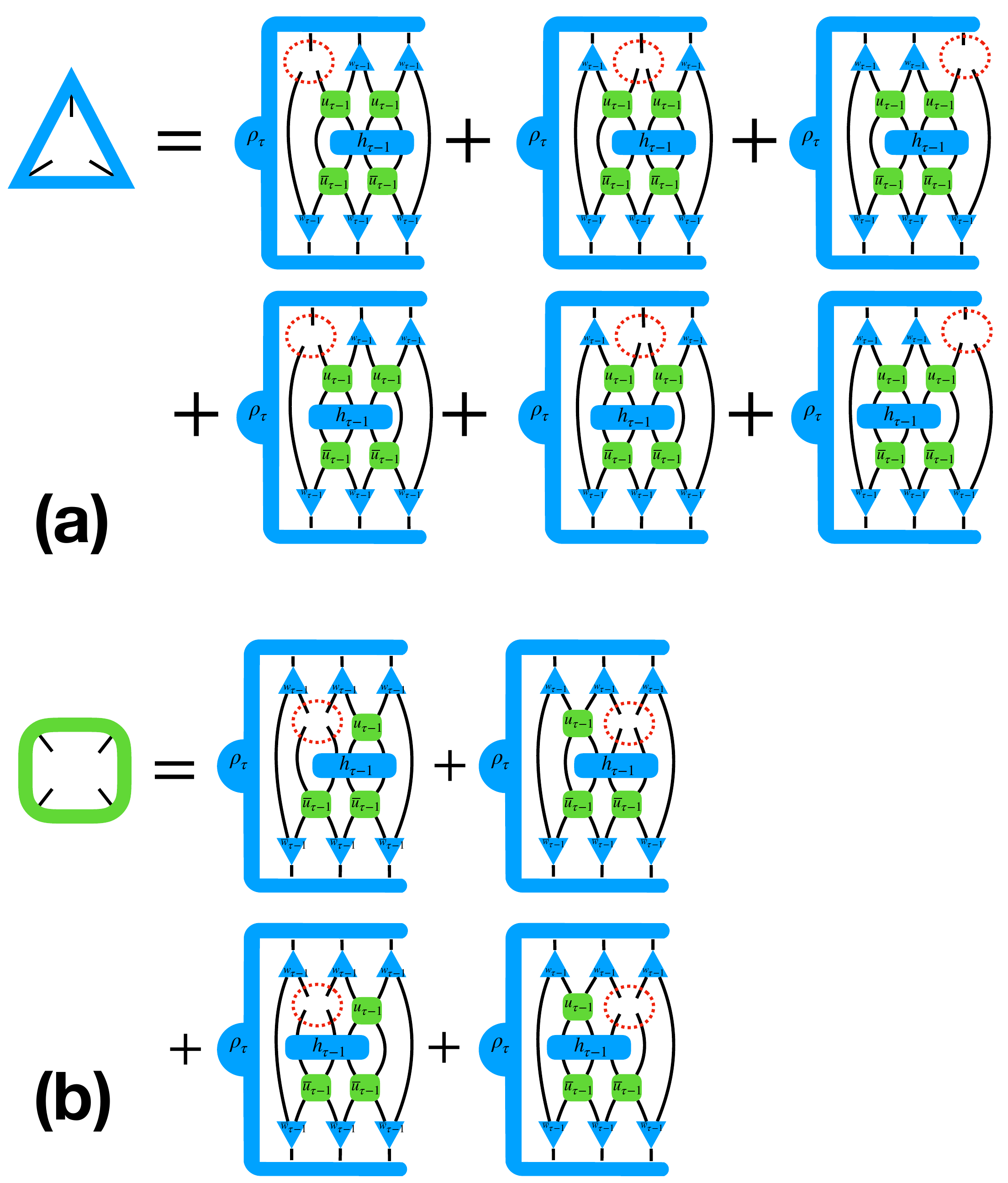}
  \caption{Contributions to the environment of the isometry $w$ (a) and
    disentangler $u$ (b). Note that these can be conveniently computed using the autodiff feature of TensorFlow.}\label{fig:all_envs}  
\end{figure*}

\section{Appendix}
\subsection{Environment contributions}
For completeness, in \Fig{fig:all_envs} we show all contributions to the environments of
isometry $w$ and disentangler $u$.

\end{document}